\documentstyle[aps,prb,floats,preprint,eqsecnum,tighten,epsfig]{revtex}
\newcommand {\beq} {\begin{equation}}
\newcommand \bea {\begin{eqnarray} \nonumber }
\newcommand{\eeq}{\end{equation}}
\newcommand \eea {\end{eqnarray}}
\newcommand{\beqa}{\begin{eqnarray}}
\newcommand{\eeqa}{\end{eqnarray} }

\newcommand{\nn}{\nonumber\\}

\begin{document}                

\begin{titlepage}
\vskip .27in
\begin{center}
{\large \bf
Mean-field theory of temperature cycling experiments in spin-glasses
}

\vskip .27in

Leticia F. Cugliandolo
\vskip .2in
{\it Laboratoire de Physique Th\'eorique de l'\'Ecole Normale Sup\'erieure de 
Paris
\\
24, rue Lhomond, 75231, Paris, France} \\
and
\\
{\it Laboratoire de Physique Th\'eorique et Hautes \'Energies, Jussieu,
\\
4, Place Jussieu, 75005, Paris, France
}
\\
leticia@physique.ens.fr

\vskip .2in

Jorge Kurchan
\vskip .2in
{\it Laboratoire de Physique de l'\'Ecole Normale Sup\'erieure de 
Lyon \\
46, All\'ee d'Italie, Lyon, France } 
\\
Jorge.Kurchan@enslapp.ens-lyon.fr

\end{center}
\date\today

\vskip 8pt

\vspace{0.5cm}
\begin{center}
{\bf Abstract}
\end{center}

 We study  analytically the effect of temperature cyclings in
 mean-field spin-glasses. 
 In accordance with real experiments, we
obtain a strong reinitialization of the 
dynamics on decreasing the temperature
combined with memory effects when 
the original high temperature is restored.
The same calculation applied to mean-field models of structural glasses
shows no such reinitialization,
again in accordance with experiments.
In this context, we derive some  relations between experimentally 
accessible  quantities and propose new experimental protocols.
Finally, we briefly discuss the effect of field cyclings during 
isothermal aging.

\vspace{2cm}
LPT-ENS/9842; LPTHE/9853.

\end{titlepage}

Glasses are  characterized by having extremely slow relaxations and
by the strong dependence of their behaviour upon the (``waiting'') 
time elapsed since their preparation. 
The latter property is usually called {\it physical aging}.

A means to study the dynamics in the glassy phase in more 
detail consists in following the evolution of the sample under 
a complicated temperature history. 
The protocols that have been more commonly used include 
temperature cyclings within the low temperature phase.

The results for different types of glasses are quite 
different.
\cite{Struick,Revihaoc,tempcycles1,tempcycles2,suecos1,saclay,suecosnew,Levelut,Sid}
{\em Spin}-glasses show the puzzling phenomenon of reinitialization of 
aging following a decrease in temperature, combined with the recall  
of the situation attained before the negative jump  when the 
original high temperature is restored.\cite{Revihaoc,tempcycles2}
 Remarkably, when similar protocols were applied 
to {\em structural} glasses, e.g. in 
dielectric constant measurements of glycerol by Leheny and Nagel,\cite{Sid}
 no substantial reinitialization was observed.\cite{foot}

This difference in the  effect of temperature changes  on 
spin and structural glasses is  a fact that any generic theory of glasses 
is expected to explain. 

\vspace{.3cm}

Different groups interpreted 
the behaviour of spin-glasses under  temperature cyclings during aging 
 as evidence for both the
droplet\cite{suecos1,suecosnew} and the hierarchical\cite{saclay} pictures
of the dynamics (the former with 
 some extra refinement
\cite{suecosnew} with respect to the original 
versions of the eighties\cite{droplets}). 

The hierarchical dynamic picture\cite{Do,saclay}
 is a heuristic way to think about the results from positive and 
negative cyclings inspired in the 
organization of equilibrium states in the Parisi solution 
of mean-field spin-glasses, such as the Sherrington-Kirkpatrick model.
It is  assumed that spin-glasses have a large number of 
metastable states that  are  organized
 in a hierarchical fashion just like the equilibrium  states.
It is then proposed that 
the system is composed of (independent) subsystems 
whose dynamics is given by the wandering in such a landscape. 
An average over subsystems has to be invoked in order to obtain
smooth results as observed in experiments.

A concrete realisation of a hierarchical dynamic system 
can be made with the trap models.\cite{Siho_old,Siho_new,traps} 
These models have been solved analytically in isothermal 
conditions.\cite{traps} Though a full analytic description of their dynamics 
in the presence of temperature cyclings is not available yet, 
a careful discussion of their effects yields  very 
encouraging  results.\cite{traps}

Surprisingly enough, the main features of
the cycling experiments have  never been derived analytically from 
 microscopic models, 
while the numerical evidence\cite{Ma,Ri,Ba,japon} is  inconclusive.
In this paper we shall show analytically how these
 effects arise in mean-field models of spin glasses, 
and why they are absent in mean field 
models of structural glasses. 
One of the questions that will receive  a clear answer is
why the effects should be hardly observable at very short times, 
such as are inevitably involved in simulations.

\vspace{.3cm}

We shall consider in detail the particular  class of temperature 
cycling experiments
in which the thermoremanent magnetization (TRM) is measured.
Similar conclusions have been extracted from 
the out of phase susceptibility ($\chi''$) data  at fixed 
frequency.\cite{Revihaoc,sitges} 

There is however a slight difference between TRM and ac $\chi''$ measurements.
In the former the TRM {\it after} the cycling is recorded and, 
since measurements are directly performed in the time domain, one has 
access to very large time scales after $t_w$. 
In the latter case the $\chi''$
is measured {\it during and after} the cycling
This allows us, for example, 
to clearly see the large reinitialization of the dynamics 
provoked by the negative jump.\cite{Revihaoc} The price to pay is that 
in ac measurements the frequencies are necessarily small compared to 
to the inverse of the measuring time. One then has access to relatively small 
time-differences.

\begin{figure}
 \centerline{\hbox{
   \epsfig{figure=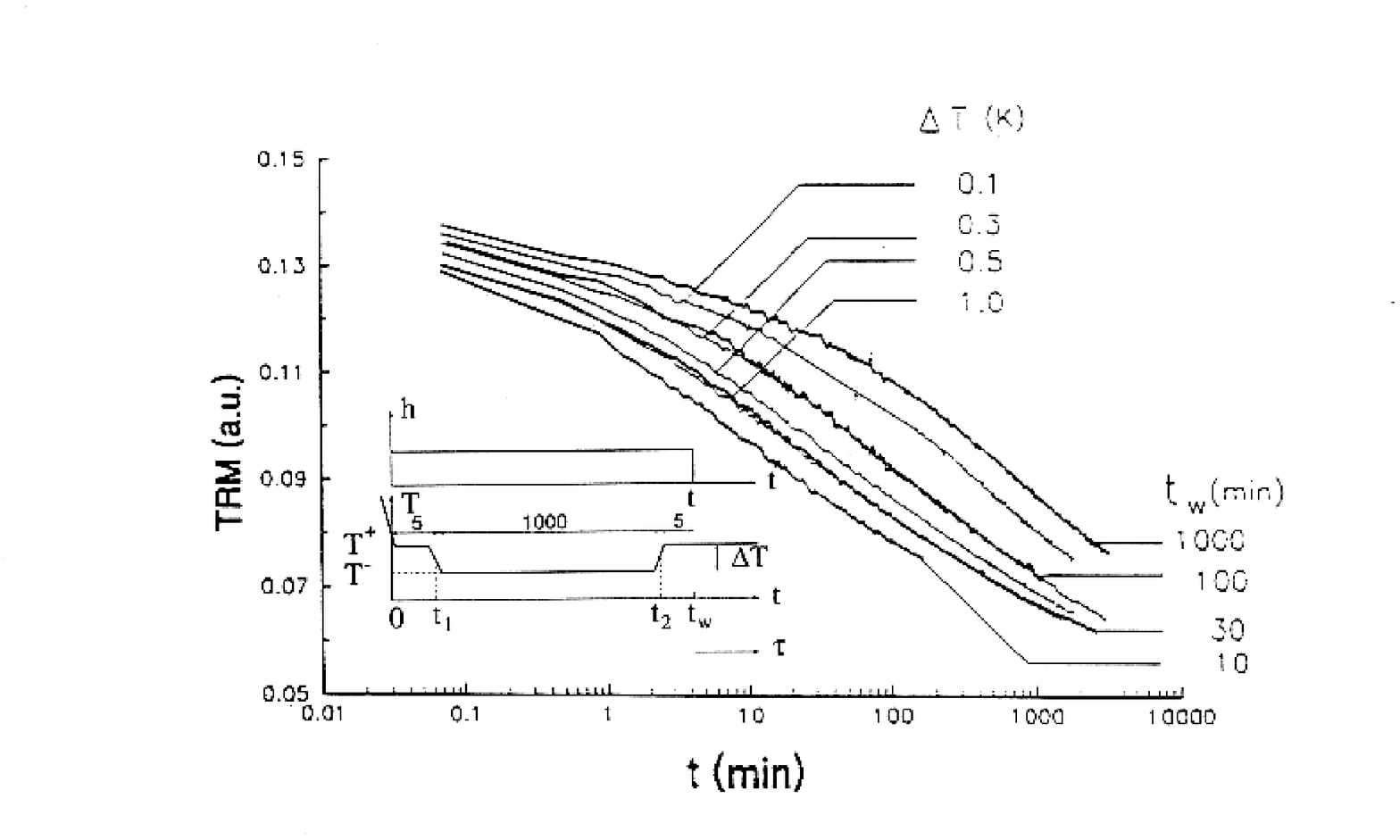,width=13cm}}
 }
 \caption{The TRM curves after a long negative cycle sketched in the inset.
The temperature at which the relaxation is measured is 
$T=12K=0.72 \, T_g$. 
The bold lines are reference curves corresponding to waiting at
a constant temperature of $12 K$ during $t_w=10,\,30,\,100,\,1000$
min. The thin lines are obtained after a temperature cycling
of $1000$ min and temperature jumps $\Delta T=0.1, \,0.3,\,0.5,\,1\, K$.
The data is courtesy of E. Vincent, see Ref.~[2].}
 \label{tempdown}
 \end{figure}

In the TRM experiments of Refregier et al.\cite{Revihaoc}
the system is quenched to a subcritical temperature
under a small field that is used as a probe, for which the linearity 
in the  response is checked within the same experiment.

In the negative temperature cycling experiment  (see the inset of 
Fig.~\ref{tempdown}),
the system is quenched at $t=0$ to a  temperature
$T^+$. At a time $t_1$ the temperature is dropped to $T^-$ and 
at a later time $t_2$ the temperature $T^+$ is restored.
The fraction of time spent at $T^-$ is rather large.

The resulting state of the sample after the temperature cycling is 
investigated by cutting off the field 
at a  time $t_w$ soon after $t_2$ and by recording the subsequent decay 
of the TRM.
The results for different $\Delta T$ are  shown in Fig.~\ref{tempdown}:
The  TRM curves for each $\Delta T$ can be superposed to the TRM
curves obtained at constant $T^+$  but with an  effective
waiting time $t_1+t_w-t_2 \leq t_w^{\sc eff} (\Delta T,t_1,t_2,t_w) \leq t_w$. 
Note that the first inequality implies that the system remembers the evolution 
performed  at the higher temperature $T^+$ while the 
time spent at the lower temperature $T^-$ is partially (or even totally)
erased. 

Once the negative cycle is over,
{\em its main effect is to slow down the aging process}. 
This result is very  intuitive and will hold for almost any system
with slow dynamics activated by thermal noise.

\begin{figure}
 \centerline{\hbox{
   \epsfig{figure=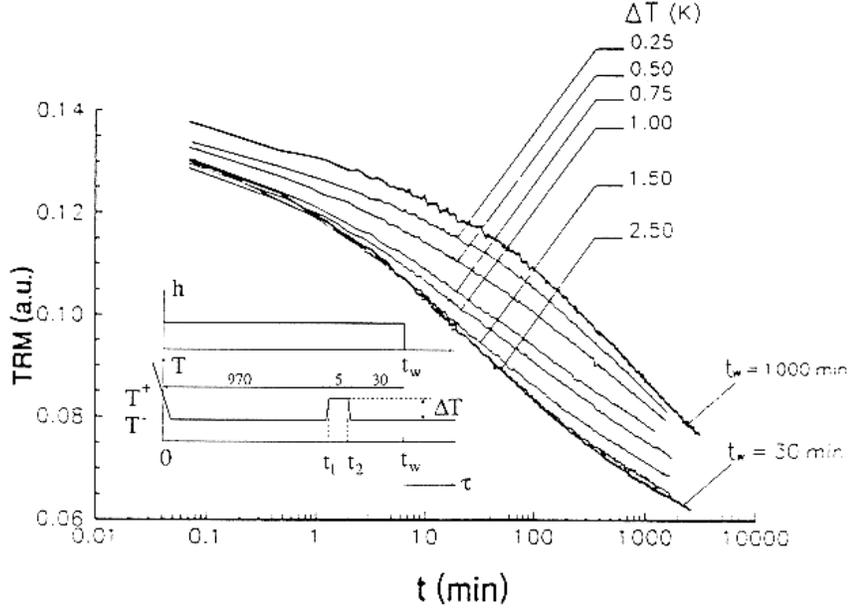,width=13cm}}
 }
 \caption{The TRM curves after a short positive cycle
sketched in the inset.
The relaxation is obtained at $T=12K=0.72\, T_g$. The bold lines are reference
curves obtained after isothermal waiting at $12K$ for $t_w=30, \, 1000$ min. 
The thin lines are the decay of the TRM
after a cycle of $5$ min and $\Delta T=0.25,\,0.5,\,0.75,\,1,\,1.5,\,2.5\, K$. 
Data is courtesy of E. Vincent, see Ref.~[2].}
 \label{tempup}
 \end{figure}

The real surprise appears when a cycle  of positive temperature 
(inset of Fig.~\ref{tempup})
is applied to spin glasses.
 Here the procedure is the inverse: 
the system is quenched to a temperature $T^-$ up to a long  time 
$t_1$. This is followed by a short period at temperature  $T^+$ 
from $t_1$ to  $t_2$,  at which  $T^-$ is restored.
As before, the field is  cut off 
at a later time $t_w$ and the  subsequent decay 
of the TRM is recorded. 
 
The result is shown in Fig.~\ref{tempup}. The higher the upward pulse
in temperature, the  generally younger the system seems but, unlike in the
negative cycling case,
{\em the effect  cannot be described  with an effective waiting time}.

Note that in the simplest cases  of aging through activated processes
(as  for example   the coarsening of the random field 
Ising model) the effect would be the reverse: the temperature pulse would 
quicken the  activated processes and help aging. Refreshment would arise only
if the temperature
pulse is high enough to take the system above the transition.

\vspace{.3cm}

Indeed, the solution for mean-field spin-glasses discussed
below involves an {\em additive } separation of the TRM curve 
$M^{\sc trm}=M^{\sc rep} + M^{\sc f}$. 
In the positive temperature cycling
the ``slow'' component $M^{\sc rep}$ is not affected by the pulse
while the  ``fast'' component
$M^{\sc f}$ is taken to its effective critical temperature and is 
thus completely
reinitialised. The amplitude of $M^{\sc f}$ ($M^{\sc rep}$) is 
larger (smaller) for higher  (lower) temperature pulses. 
In the negative temperature cycling one is essentially probing only
the slow component $M^{\sc rep}$ that is slowed down by the effect
of the lower temperature excursion. 

\vspace{.3cm}

We shall show that the difference in the  effects of positive and negative 
temperature changes is present in the mean-field version of
spin-glass models while it is absent in the models 
thought to be a mean-field caricature of structural glasses.
The relevant difference between these models resides in 
their dynamic behaviour below the transition. The former decay
in infinitely many time-scales while the latter do in only 
two. (For an unambiguous definition of 
time-scales in aging problems see 
Ref.~[\raisebox{-.22cm}{\Large \cite{Cuku2}}].) The explanation we 
give here is based on this difference. We choose to discuss in 
detail the magnetization behaviour although 
the susceptibility data, in particular the reinitialization after a negative 
jump, can also be understood within this framework. 

Another common way to test the dynamics in the glassy phase is to apply 
magnetic field steps during aging at constant temperature. 
In these experiments no important difference is 
obtained between switching on and switching off the dc field. We shall 
briefly discuss this result within the same analytic framework. 

The organisation of the paper is the following. In Section~\ref{section1}
the main features of the 
analytic approach are described. In Section~\ref{section2} the 
temperature cycling experiments 
are explained within this analytic approach. 
The results of field cycling experiments are confronted to this approach in 
Section~\ref{section3}. Section~\ref{section4} is devoted to some new
experimental proposals and Section~\ref{section5} to the conclusions.

\section{Theoretical approach}
\label{section1}

In order to understand the effect of temperature cyclings on
the relaxation of the TRM and out of phase susceptibility 
of mean-field spin-glass models we have to understand the 
time-dependence of the auto-correlation and response functions:
\beq
C(t,t') \equiv 
\frac1N \, \sum_{i=1}^N \langle \overline{ s_i(t) s_i(t') } \rangle
\; ,
\;\;\;\;\;\;\;\;\;\;\;
R(t,t') \equiv 
\left.
\frac1N \, \sum_{i=1}^N \frac{\delta \langle \overline{ s_i(t) } \rangle}
{\delta h_i(t')}
\right|_{h=0}
\; .
\label{gg}
\eeq
For mean-field models, one obtains 
a set of coupled equations that entirely determine the dynamics of 
these two-point functions. 
The thermoremanent magnetization $M^{\sc trm}(t,t_w)$ at time $t$, after 
cutting off a small field at time $t_w$, is then 
expressed in terms of the integrated
susceptibility $\chi(t,t_w)$:
\beq
\chi(t,t_w) = \int_{t_w}^t d\tau \, R(t,\tau) \;,
\;\;\;\;\; M^{\sc trm}(t,t_w)=h[\chi(t,0)-\chi(t,t_w)]
\eeq
with $h$ the strength of the small dc field applied.

\subsection{The model}

For definiteness let us consider the toy model consisting of $N$ 
continuous spins with a spherical constraint $\sum_i^N s_i^2(t)=N$ and a random energy $V({\bbox s})$ 
correlated as $\overline{V({\bbox s})V({\bbox s'})}=N\nu({\bbox s} \cdot {\bbox s'}/N)$.
The statics\cite{Mepa} as well as the constant temperature dynamics of 
this model have been solved in all detail 
(see Refs.~[\raisebox{-.22cm}{\Large \cite{Cuku,Frme,review}}]).
Two types of models with  potential correlations such that  
$1/\sqrt{\nu''(C)}$ is, for all $0 \leq C \leq 1$, concave (convex) 
yield very different statics\cite{Mepa} 
(one step replica symmetry breaking versus 
full replica symmetry breaking) and dynamics\cite{Frme}
(first order versus second order dynamical transitions) and
correspond to mean-field versions of  
structural and spin glasses respectively. 
Examples that have been extensively studied in the literature 
are the ``$p=3$ spherical model'' for glasses,\cite{Cuku} 
with $\nu(C)=1/2 C^3$ (and $(\nu''(C))^{-1/2}$ concave) and 
the ``$p=2$ plus $p=4$ model'' for spin-glasses,\cite{Theo} 
with $\nu(C)=1/2(c_1 C^2+c_2 C^4)$ 
and $c_1$ and $c_2$ two constants such that
$(\nu''(C))^{-1/2}$ be convex.

The exact equations of motion for $C$ and $R$ at times 
$t > t'$ are \cite{review}
\begin{eqnarray} 
\frac{\partial C(t,t') }{ \partial t} 
&=&
 -z(t) \, C(t,t') +\int_0^{t'} d\tau \, D[C(t,\tau)] \, R(t',\tau)
+ \int_0^t d\tau \, \Sigma[C(t,\tau)] \, C(\tau,t')
\; ,
\label{eq21a}
\\
\frac{\partial R(t,t')  }{ \partial t} 
&=&
 -z(t) \, R(t,t') +  
   \int_{t'}^t d\tau \, \Sigma [C(t,\tau)] \, R(\tau,t')
\; ,
\label{dyneq}
\\
z(t)&=&
T(t) + 
\int_0^t d\tau \; \left( D[C(t,\tau)] R(t,\tau) + \Sigma[C(t,\tau)] C(t,\tau) \right)
\; ,
\label{zeq}
\end{eqnarray}
with $\Sigma(t,\tau) \equiv  D'[C(t,\tau)] R(t,\tau)$, 
$D'[C]=\partial_C D[C]$ and $D[C]=\nu'(C)$. The function $z(t)$ is a 
Lagrange multiplier that enforces the spherical constraint.
We shall concentrate on the spin glass-like case
that corresponds to a concave random energy correlation. 
We shall briefly discuss the structural glass-like case
at the end of this Section and explain why it does not 
show large asymmetry effects.
Let us recall the constant temperature solution
for the mean-field spin-glass models.

\vspace{.3cm}

One of the salient features of the relaxation of mean-field 
spin-glasses below the dynamic critical temperature 
$T_d$ is the presence of infinitely many (two-time) 
scales organised in a hierarchical way. 
In the low temperature phase the 
two-point correlation and response depend on the 
two times involved. The form of the relaxation is usually
analyzed by looking at these functions at fixed (but large) waiting time
in terms of the time difference $\tau=t-t_w$. The correlation and response
functions have a first fast stationary relaxation; for instance, 
the correlation rapidly
decays from $1$ to $q_{EA}$,  the 
Edwards-Anderson parameter. This time-scale is usually called the FDT
regime, for reasons that will become clear below.
For longer time-differences the relaxation continues at a 
waiting-time dependent speed. Furthermore, 
if one imagines the subsequent decay of the correlation
as taking place in infinitesimal steps,
each step takes much longer than the previous one. 
Indeed, in the limit of large waiting time each infinitesimal step
implies a period of time that is infinitely longer 
than the previous one and 
these time-scales get completely separated.  The same separation of time scales
characterizes the decay of the response.

The sharp separation of time-scales allows us 
to split the correlations in a fast part $C^{\sc f}(t,t')$, 
going from one at equal times to $q_0$ at
very distant times, and a slow part $C_{q_0}(t,t')$, 
going from $C_{q_0}(t,t)=q_0$ and tending
to zero at even more distant times. The point at which we 
split the correlation {\it is chosen arbitrarily} provided it satisfies
$q_0 \leq q_{EA}$.
Correspondingly we separate the response
in a fast and a slow part. We then have
\beqa
C(t,t') &=& C_{q_0} (t,t') + C^{\sc f}(t,t') - q_0 \; ,
\\
R(t,t') &=& R_{q_0} (t,t') + R^{\sc f}(t,t')
\; .
\label{separation}
\eeqa
Since $C_{q_0}$ ($R_{q_0}$) is infinitely slower than $C^{\sc f}$
($R^{\sc f}$) their time evolution can be characterized as follows: 
for all times such that $C^{\sc f}$ changes
$C_{q_0}$ is just constant and equal to $q_0$. Instead, in the time-regime 
in which $C_{q_0}$ varies $C^{\sc f}$ has achieved its asymptotic
 value $q_0$ and does not further evolve.

Note that the time-scale separation 
is achieved {\it only} in the large waiting-time limit. 
This is the crucial ingredient for the argument we shall 
use to show that Eqs.~(\ref{eq21a})-(\ref{zeq}) capture the 
phenomenology of temperature cyclings. 
We believe that the effects disussed in the Introduction 
have not been clearly 
observed numerically because the times explored were inevitably very 
short and the time-scales could not be sufficiently separated.\cite{Ma,Ri,Ba}

As we shall show, the change in time-dependence of the fast and slow 
parts of $C$ and $R$ 
under a temperature jump are very different. The slow 
parts $C_{q_0}$ and $R_{q_0}$ are modified through a 
smooth time-reparametrization. The fast parts $C^{\sc f}$
and $R^{\sc f}$ behave as the correlation and response of 
an effective model with a critical temperature precisely
equal to $T^+$. The temperature jumps have a  strong effect 
on the fast parts and
this effect is very different depending on the sign of the change.

\vspace{.3cm}

The other salient feature of the  low-temperature dynamics of mean-field
spin-glasses is the violation of the 
fluctuation-dissipation theorem (FDT) and, most 
importantly, the generalized form that the relation
between correlation and response takes.   

A very useful way to quantify the fluctuation-dissipation 
relation that holds out of equilibrium is 
given by the relation between
the integrated response $\chi(t,t_w)$  and the correlation
$C(t,t_w)$.\cite{Cuku2} At fixed and large $t_w$, one constructs a plot
of $\chi(t,t_w)$ vs $C(t,t_w)$ using $t$ as a parameter. 
This is shown in Fig.~\ref{chifig} for two temperatures $T^+$ and $T^-$.
For each temperature the $\chi$ vs. $C$ curve consists of two parts:

\begin{figure}
\centerline{\hbox{
\epsfig{figure=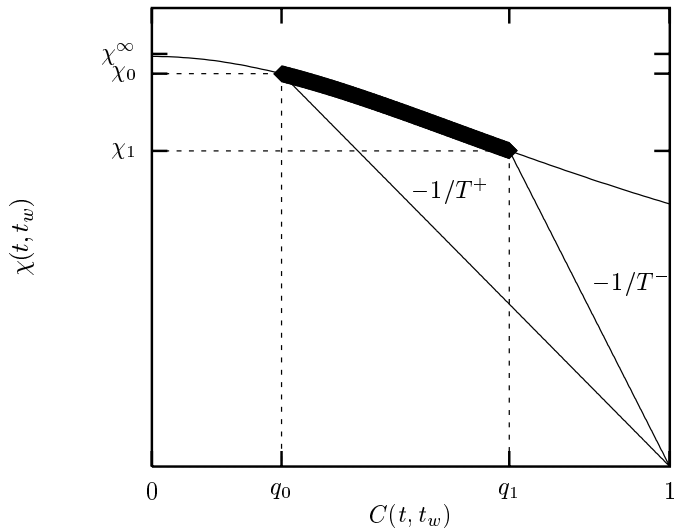,width=11cm}}
}
\vspace{-.5cm}   
\caption{Sketch of the $\chi$ vs $C$ plot. The two straight lines have 
slopes $-1/T^+$ and $-1/T^-$ and correspond to the FDT result. 
The curve is temperature independent and it is given by $(\nu''(C))^{-1/2}$. 
The bold segment 
is the part erased (created) when decreasing (increasing) the temperature. 
See text for more details.}
\label{chifig}
\vspace{-.1cm}
\end{figure}

\begin{itemize}
\item
A straight line of slope minus the inverse temperature. This corresponds
to the fast time regime where the FDT
holds and $C$ decays from $1$ to $q_{EA}$. We call this time-regime the FDT 
regime.
\item
A curve given by
\begin{eqnarray}
f(C) 
&=&
\frac{1}{\sqrt{\nu''(C)}}
\; .
\label{chicurve}
\end{eqnarray}
This corresponds to the slower time-regimes where the FDT
is violated and $C$ decays from $q_{EA}$ to $0$ in a waiting-time 
dependent manner.
\end{itemize}

For temperature $T^+$  the $\chi$ vs $C$ plot follows the  straight line  
of gradient $-1/T^+$ from
$(1,0)$ to $(q_0,\chi_0)$ and then the curve $f(C)$ 
up to $(0,\chi^\infty)$. For temperature
 $T^-$ it is given by  the  straight line  of gradient $-1/T^-$ from
$(1,0)$ to $(q_1,\chi_1)$ and then the curve $f(C)$ up to $(0,\chi^\infty)$.
The Edwards-Anderson parameters are $q_1$ and $q_0$, respectively. 
The two extreme cases are $T=0$ and $T=T_c$ (the critical temperature).
The former corresponds to a vertical line starting at $(1,0)$ that matches 
the curve $f(C)$ at $C=1$ and then follows it up to 
$(0,\chi^\infty)$. The $\chi$ vs $C$ plot for the critical temperature
is just a straight line linking $(1,0)$ and $(0,\chi^\infty)$ with a slope
$-1/T_c$.

The remarkable fact of this family of models is that the curved 
segment  from $(q_0,\chi_0)$  to $(0,\chi^\infty)$ is the same for
the temperatures below $T^+$. 
This is not a general property of mean-field
spin-glass models. It holds exactly for the family of models here considered
but only approximately for the Sherrington-Kirpatrick model.
In the static replica approach, the corresponding property of 
the function $x(q)$ is called the ``Parisi-Toulouse approximation''.\cite{Pato}
The numerical evidence seems to show that this approximation works
very well in finite dimensional spin-glasses,\cite{Roma} 
at least within numerically accessible times.

The analytic argument we shall develop
below can be most easily pictured by considering Fig.~\ref{chifig}.
The range of correlation and response values in which the system
has aging dynamics is given by the curved part of the plot.
{\em On changing the temperature, the straight line corresponding
 to the fast relaxation
moves clockwise and anticlockwise  like a  windshield-wiper,
 creating and destroying
the bold segment $(q_0,\chi_0) \rightarrow (q_1,\chi_1)$, thus restarting and
erasing the aging scales corresponding to this interval.}

\subsection{Analysis}

In the limit of large waiting times
we can separate the different time-regimes
as follows. The equations for the {\it fast parts} of the  decay are 
\begin{eqnarray} 
\frac{\partial C^{\sc f}(t,t') }{ \partial t} 
&=&
 - (z^{\sc f}  (t)+ {\bar z}(t))  \, C^{\sc f}(t,t') 
  +\int_0^{t'} d\tau \;  D[C^{\sc f}(t,\tau)] \; R^{\sc f}(t',\tau)
\nonumber\\
& & 
+ \int_0^t d\tau \;  \; D' [C^{\sc f}(t,\tau)] \;  R^{\sc f}(t,\tau) C^{\sc f}(\tau,t') +\bar z(t)
\; ,
\label{eq21aa}
\\
\frac{\partial R^{\sc f}(t,t')  }{ \partial t} 
&=&
 - \, (z^{\sc f} (t)+ {\bar z}(t))R^{\sc f}(t,t') +  
   \int_{t'}^t d\tau \;
\; D' [C^{\sc f}(t,\tau)] R^{\sc f}(t,\tau)  \; R^{\sc f}(\tau,t')
\; ,
\\
z^{\sc f} (t)&=& T(t) + \int_0^t d\tau \left(
D[C^{\sc f}(t,\tau)]  +  D'[C^{\sc f}(t,\tau)]  C^{\sc f}(t,\tau)
\right) \, R^{\sc f}(t,\tau)
\; .
\label{dyneqq}
\end{eqnarray}  
The equations for the {\it slow parts} of the decay are
\begin{eqnarray} 
small
&=&
 -  \left (z^{\sc f}(t) + \overline z(t)-
           \int_0^t d\tau \, D'[C^{\sc f}(t,\tau)] R^{\sc f}(t,\tau) 
    \right)  
    \, C_{q_0}(t,t') 
\nn
& & 
+ \chi^{\sc f}(t') \, D[C_{q_0}(t,t')]  
  +\int_0^{t'} d\tau \,  D[C_{q_0}(t,\tau)] \, R_{q_0}(t',\tau)
\nonumber\\
& & 
+
\int_0^t d\tau \, D'[C_{q_0}(t,\tau)] \, R_{q_0}(t,\tau) \, C_{q_0}(\tau,t')
\; ,
\label{eq21ab}
\\
small
&=&
 -  \left(z^{\sc f}(t) +\overline z(t) - D'[C_{q_0}(t,t')]
\chi^{\sc f}(t') - 
\int_0^t d\tau \, D'[C^{\sc f}(t,\tau)] R^{\sc f}(t,\tau) \right)  \, R_{q_0}(t,t') 
 \nonumber\\
& & 
+ 
 \int_{t'}^t d\tau \;
D' [C_{q_0}(t,\tau)] R_{q_0}(t,\tau) \; R_{q_0}(\tau,t')
\; ,
\label{dyneqb}
\\
{\bar z}(t)&=&
\int_0^t d\tau \;
 \left( 
D[C_{q_0}(t,\tau)] + D'[C_{q_0}(t,\tau)] C_{q_0}(t,\tau) 
\right) \, 
R_{q_0}(t,\tau)
\; .
\label{coci}
\end{eqnarray}  
The input of the slow scale on the fast one is given by the 
 one-time quantity ${\bar z}(t)$. 
The fast scale influences the slow one through three 
factors: 
$z^{\sc f}(t)$ defined in Eq.~(\ref{dyneqq}), 
$\chi^{\sc f}(t)$ and $\tilde \chi^{\sc f}(t)$ 
defined as
\begin{eqnarray}
\chi^{\sc f}(t)&=&\int_0^t d\tau \,  R^{\sc f}(t,\tau)
\; ,
\nn
\tilde \chi^{\sc f}(t) &=& \int_t^\infty \, d\tau R^{\sc f}(\tau,t)
\; ,
\end{eqnarray}
and the last term in the parenthesis of Eqs.~(\ref{eq21ab}) and 
(\ref{dyneqb}). One can eliminate the dependence on this last factor 
by using the relation 
\beqa
small \sim 
-
\left (z^{\sc f}(t) + \overline z(t)-
           \int_0^t d\tau \, D'[C^{\sc f}(t,\tau)] R^{\sc f}(t,\tau) 
    \right)  
    \, q_0
+ \chi^{\sc f}(t) \, D[q_0]  
+\overline z(t)
\label{relation}
\eeqa
that follows from Eq.~(\ref{eq21ab}) evaluated at
\beq
t' \to t \;\;\;\;\;\;\;\;
\Rightarrow \;\;\;\;\;\;\;\;
C_{q_0}(t,t') \to q_0
\; .
\eeq
After inserting the relation (\ref{relation}) in
Eqs.~(\ref{eq21ab}) and (\ref{dyneqb})
one has the following set of slow equations
\beqa
small
&=&
 - \frac{1}{q_0} (\overline z(t)+\chi^{\sc f}(t)D[q_0])  \, C_{q_0}(t,t') 
  +\int_0^{t'} d\tau \;  D[C_{q_0}(t,\tau)] \; R_{q_0}(t',\tau)
\nonumber\\
& & 
+ \chi^{\sc f}(t') D[C_{q_0}(t,t')]
+
\int_0^t d\tau \;  \; D'[C_{q_0}(t,\tau)] R_{q_0}(t,\tau) \; C_{q_0}(\tau,t')
\; ,
\label{eq21ab2}
\\
small
&=&
 - \,\frac{1}{q_0} (\overline z(t)+\chi^{\sc f}(t)D[q_0])  R_{q_0}(t,t') +  
\tilde chi^{\sc f}(t') D'[C_{q_0}(t,t')] R_{q_0}(t,t')
 \nonumber\\
& & 
+ 
 \int_{t'}^t d\tau \;
D' [C_{q_0}(t,\tau)] R_{q_0}(t,\tau) \; R_{q_0}(\tau,t')
\; ,
\label{dyneqb2}
\eeqa
together with Eq.~(\ref{coci}) for $\overline z(t)$. 
The effect of the fast scales on the slow one is then only given 
by $z^{\sc f}(t)$, $\chi^{\sc f}(t)$ and $\tilde \chi^{\sc f}(t)$.

At constant temperature  the quantities 
${\bar z}(t)$,  $\chi^{\sc f}(t)$ and $\tilde \chi^{\sc f}(t)$ tend to
the following limits
\begin{equation}
{\bar z}(t) \rightarrow  {\bar z}^\infty 
\; , 
\;\;\;\;\;\;\;\;
\chi^{\sc f}(t)\rightarrow  \chi_0
\; , 
\;\;\;\;\;\;\;\;
\tilde \chi^{\sc f}(t)\rightarrow  \chi_0
\; .
\end{equation}
As a consequence of the independence of temperature 
of $\chi(C)$ if $C < q_{EA}$, see Fig.~\ref{chifig},
{\em neither ${\bar z}^\infty$ nor $\chi_0$ depend on the temperature}.

\vspace{.3cm}

Consider now the temperature cycling experiment. 
We consider the effect on the slow and fast parts
separately and denote the correlation after the cycle 
\beq
C^{\sc cycle}(t,t') = C^{\sc rep}_{q_0}(t,t') + 
\hat C^{\sc f} (t, t')
\; .
\label{change}
\eeq
The first term is the slow part while the 
second is the fast part. Similarly, for the TRM:
\beq
M^{\sc cycle}(t,t') = M^{\sc rep}_{q_0}(t,t') + 
\hat M^{\sc f} (t, t')
\; .
\label{change1}
\eeq

Under the assumption that in the presence of the cycling
$C_{q_0}$ and $R_{q_0}$
are  still slower than $C^{\sc f}$ and $R^{\sc f}$ and are affected at most by
time-reparametrisations, an assumption whose consistency has to be checked,
${\bar z}(t)$ remains constant and equal to ${\bar z}^\infty$
throughout the cycle. 
The function $\chi^{\sc f}(t)$ stabilises to the temperature-independent value 
$\chi_0$ soon after each temperature change on a time-scale that 
is very short compared to any aging times since it is the relaxation of a 
one time-quantity.\cite{numerical}
Apart from a short pulse, proportional to  $\chi(t)-\chi_0$, the slow
equation feels the effect of the temperature cycling {\em only through 
the small matching terms}, that we collected on the left-hand-side of Eqs.~(\ref{eq21ab}) and 
(\ref{dyneqb}) under the name of {\it ``small''}.
These affect only reparametrizations,\cite{review} consistently
with the initial assumption in the discussion.

Hence, the {\it slow parts} of the correlation and response 
are modified by the cycling through a smooth
change in the speed of evolution, i.e. a {\it time-reparametrisation}.
This is more precisely stated by studying the change of
the slow parts in each time-scale. With this purpose, let us 
denote $C_\alpha^{T^+}(t,t')$ and $R_\alpha^{T^+} (t,t')$ 
the correlation and response function in the time-scale
labelled $\alpha$ for a system aging at constant temperature $T^+$.  
In the presence of a temperature cycle between $t_1$ and $t_2$ 
each of these is modified according to
\begin{eqnarray*}
C^{\sc cycle}_\alpha (t,t')= C_\alpha^{T^+} (h_\alpha(t),h_\alpha(t')) 
\; ,\;\;\;\;\;\; 
R^{\sc cycle}_\alpha (t,t')= \partial_{t'}{h_\alpha} (t') \; 
R_\alpha^{T^+} (h_\alpha(t),h_\alpha(t'))
\; .
\end{eqnarray*}
Though the calculation of $h_\alpha(t)$ is beyond our analytical means,
we know that it has three distinct behaviours depending on the relation 
between $t'$, $t$ and $t_1$, $t_2$. Obviously,
$h(t) = t$ if $t<t_1$; $h_\alpha(t)$ has inflections around 
$t_1$ and $t_2$; and 
we expect $\partial_t h_\alpha(t) \sim 1$ for $t \gg t_2$ when the system 
forgot the cycle, in accordance with the {\it weak long-term memory} 
property.\cite{Cuku,review}
Furthermore, we expect that during
the period $t_1<t<t_2$
the dynamics slows down, $\partial_t h_\alpha(t)<1$,
if we cool down the sample,  
while the dynamics accelerates, $\partial_t h_\alpha(t)>1$, 
if we heat the sample.
In Eq.~(\ref{change}) we have generically denoted the reparametrized slow part 
$C_{q_0}^{\sc rep}$.

The fast equations (\ref{eq21aa})-(\ref{dyneqq}) decouple from the slow 
equation
and correspond to an effective model in a constant field, since
one can check that  the (essentially constant) term 
$\overline z(t)$ that appears in the fast equation can be 
looked upon as the effect of an effective magnetic field.
In fact, $T^+$ is
the critical temperature of the  model in the presence of such a field.

The temperature of the effective fast model is cycled 
following any of the two experimental protocols between its subcritical 
temperature $T^-$ and its critical temperature $T^+$.
This is why 
the effect of the temperature change is stronger on the 
fast part of the correlations and heavily depends on the sign of the change.

\vspace{.3cm}

The structural-glass models in which $(\nu''(C))^{-1/2}$ is concave
have a quite simpler isothermal dynamic behaviour below its
transition. They decay in only two two-time scales with a first stationary
decay towards $q_{EA}$
and a subsequent slower non-stationary decay towards zero. 
For these models
we are not allowed to separate the correlation and response in 
the additive way proposed in Eq.~(\ref{separation}) for any $q_0$ 
and all the analysis
presented in this seccion simply does not apply.

\section{Comparison with the experimental results}
\label{section2}

Let us now discuss the experimental TRM curves in the light of these results. 

\subsection{Negative temperature cycling}

The straight line in the $\chi$ vs.~$C$ curve presented in 
Fig.~\ref{chifig}
moves  clockwise and then anti-clockwise to its original position.
At the end of the cycle 
the bold part of the curve  disappeared from 
the $\chi$ vs $C$ plot.
At $t=0$ the fast model is quenched down to its critical temperature $T^+$,
and rapidly equilibrates. 
At time $t_1$ it is further quenched to its subcritical temperature $T^-$
and starts aging with the 
effective waiting time $t_w^{\sc eff}=t_w-t_1$. 
Finally, at time $t_2$ the fast model  is brought again to the critical 
temperature $T^+$, and it rapidly reequilibrates. 
The fast part
has only made an excursion into its glassy phase through its quench to 
$T^-$ but, since it has been taken back to its critical temperature 
$T^+$, it has quickly forgotten it.\cite{foot2} 

One concludes that there is then an effective waiting-time $t_w^{\sc eff}$
\beq
t_1 + t_w-t_2 \leq t_w^{\sc eff}(\Delta T,t_1,t_2,t_w) \leq t_w
\label{tweff}
\eeq
that characterizes the evolution of the whole system, with 
the time spent in the lower temperature 
partially contributing to $t_w^{\sc eff}(\Delta T,t_1,t_2,t_w)$.

Since the effects of  negative cooling cycles are small, 
the fraction of time spent
at $T^-$ needs to be  rather large to observe them. 
Fig.~\ref{tempdown} displays a set of curves 
measured after performing negative temperature cyclings of
different magnitude (thin  lines) and compares them to
four isothermal curves associated to smaller waiting times. 
It is clear from the figure that the curves associated to
each temperature cycling can be completely superposed to a reference 
isothermal curve (compare, for example, the 
curve for $\Delta T=T^--T^+= -0.3 K$, $t_1=15$ min, $t_2=1015$ min
and $t_w=1030$ min with the isothermal curve for $t_w=100$
min).   The system kept a {\it memory} of the evolution during $t_1$
when the temperature is raised back to its original value $T^+$.
The time spent at the low temperature has a partial effect.
There is an effective waiting time $t_w^{\sc eff}$
for the full system that verifies Eq.~(\ref{tweff}).
The effective waiting time $t_w^{\sc eff}$
decreases when one cools down the system to a lower temperature
until the effect of the time spent at the lower temperature becomes 
completely negligeable. In the experimental curves this happens for
$\Delta T=-1K$. In this case, 
the effective aging time is simply $t_w^{\sc eff}=t_1+t_w-t_2=30$ 
min.

\subsection{Positive temperature cycling}

In this case the straight line in  Fig.~\ref{chifig}
 moves first anti-clockwise and then clockwise, exposing
at the end of the cycle
the bold segment of the curve.

As already discussed, 
the slow part of the correlation is only modified by a 
time-reparametrization and, since the temperature excursion has a 
short duration, it
is hardly affected at all.

The protocol is, from the point of view of
the fast model, as follows: at time $t=0$
it is quenched to a subcritical temperature $T^-$ and let age 
until $t_1$. At this time it is taken to its critical 
temperature $T^+$ until time $t_2$,  fully reinitializing
its age. Finally, it is 
 quenched again into its glassy phase at time $t_2$ and it starts 
 aging again. It is then clear that the fast model has an
age  $t_w^{\sc eff} = t_w-t_2 < t_w$.

The observed TRM curves are then a superposition of a slow
part, which is essentially unaffected by the temperature cycle,
plus a fast part that is completely reinitialized.
The relative magnitude of fast and slow parts depends upon the
height of the positive excursion.
Thus,  for given $T^-$, 
the larger $T^+$ the faster the total decay of the TRM
as shown in Fig.~\ref{tempup} where the TRM curves for 
temperature cyclings of the same duration but different
heights are displayed.  

Having the whole decay of the TRM curves as a function of
the time-difference $t-t_w$ one can discuss them in  great detail.  
 One easily notices from the curves in Fig.~\ref{tempup} that
  one cannot associate an effective 
waiting time to the full TRM decay after the cycling, consistently with the 
the {\em additive} solution  obtained analytically
for the rejuvenation process.

Within the analytic framework, the larger the temperature 
$T^+$, the smaller the $q_0$ and $C_{q_0}$, 
 implying that the fast part of the correlation and response  
that is completely refreshed  (and hence the total rejuvenation) 
 is larger, as we see in  Fig.~\ref{tempup}. 
 
It has been stressed\cite{saclay} that
 the TRM curve after a positive temperature cycle 
 at first coincides with 
the isothermal TRM curve of  waiting time $t_w \sim t-t_2$, the difference only
becoming manifest after rather long times. 
If one compares the TRM curves in Fig.~\ref{tempup} for several 
values of $T^+$ with   
the isothermal curve for $t_w=30$ min, one observes that 
the departure from this reference curve is achieved later for
the larger $T^+$s.  

{\em This is precisely what we would expect from  the many-scales picture:}
Two protocols with the same $T^-$ and slightly different high temperatures
$T^+$ and $T^{+'}$ (associated with $q_0$ and $q_0'$) will  differ in  
the refreshment of the scales of $C$ and $M$ with  $q_0<C< q_0'$,
and these are slower the smaller  $q_0$.  Hence, the difference in TRM
between the two protocols will only show up at long times, the
higher $T^+$.

 This effect should become more and more marked for longer experiments.

\section{Field Cyclings}
\label{section3}

Field cyclings at constant temperature are known
 to yield rather large reinitializations in the out of phase 
susceptibility both on
increasing {\it and} on decreasing  the field.\cite{fieldcycles}
Although we have not done a full analysis for
the field cycling experiments, we may hint the reason why the 
arguments we used to justify a large asymmetry in temperature jump experiments
do not carry through unaltered to the field cycling case.

The effect of a time-dependent field $h(t)$ on the dynamic
equations is to add a term 
\beq
A(t,t')=h(t) \int_0^{t'} d\tau \, h(\tau) R(t',\tau)
\label{ufa}
\eeq
 to Eq.~(\ref{eq21a}), and a term 
 $A(t,t)$ to Eq.~(\ref{zeq}). 
In the constant field situation, these terms have the effect of 
eliminating the {\it slowest} scales. The $\chi$ vs~$C$ plot of Fig.~\ref{chifig} 
for a constant field $h$ coincides with the one for $h=0$ but terminates
at an $h$-dependent point $(q_0,\chi_0)$. 
The most outstanding 
effect of switching on (switching off) a field is then to erase
(create) the slowest scales between $(0,\chi^\infty)$ and $(q_0,\chi_0)$.
Note that this is already very different from the effect of temperature 
changes that mainly affect the {\it fastest} scales in the problem. 

One could then ask if a system that has been
evolving at zero field and is suddendly taken to finite $h$ does preserve 
the fast parts $C^{\sc f}$ and $R^{\sc f}$ 
without reinitialization (just the mirror image of
what an {\em increase } of temperature does).    
In fact, we can propose that such a solution exists, and see 
where the argument takes us.
We hence assume  a separation  into a fast and a slow parts of the correlation
and response just as in Eq.~(\ref{separation}) with the value $q_0$ chosen to 
be the smallest possible correlation for a field $h$.

If we now assume that on turning on the field the fast evolution remains
essentially unaltered, while the slow evolution in $C_{q_0}$ remains
slow and finally tends to $q_0$ 
(as in a constant field case), we can separate out the
fast part. The fast equations so obtained are just like 
Eqs.~(\ref{eq21aa})-(\ref{dyneqq}) with additional terms 
which (just as in the temperature cycling case
happened for the slow parts) are constant up to, and long after the field jump.
The difference is that now the ``bump'' in this otherwise constant factors and
terms is no longer short-lasted with respect to
the fast times, precisely because it comes from the slow equation and from 
the term (\ref{ufa}), which has waiting time effects itself.

Hence, the initial assumption that a decoupled equation for the fast part could
be found does not hold, at least in the manner it did for the slow parts
in the temperature cycling case. The fast parts must be altered by the 
field cycling in an important way implying no big difference between
switching on and off the field.

\section{Proposal for new experimental measurements}
\label{section4}

The obvious direct way of constructing the $\chi$ vs $C$ plot
is to perform independent measurements of the time-correlation of  
the magnetization noise $C(t,t_w)$ and the integrated response $\chi(t,t_w)$
at a given temperature and field. 
One can a posteriori check the independence on temperature and field 
of these curves.

If the assumption of independence of temperature and field of the 
$\chi$ vs $C$ curve in its aging part holds true, 
this curve can be simply obtained in a manner that is much more 
simple experimentally as it does not involve noise measurements
and only ac susceptibilities.
Indeed,  let us define the quantity $y(T,h)$
as follows 
\beq
y(T,h)\equiv
\lim_{\omega\to 0} \lim_{t_w\to\infty} \chi'(\omega, t_w) = 
\frac{q_d - q_{EA}(T)}{T}
\; .
\label{y}
\; 
\eeq
The Edwards-Anderson parameter $q_{EA}(T)$
would be then independent of the field $h$ below the 
de Almeida-Thouless line (or ``pseudo de Almeida-Thouless
line'' if it is only a dynamic crossover)
and 
\beq
y(T,h)=y(T)
\label{guess}
\; .
\eeq
The parameter $q_d$ corresponds to 
\beq
q_d \equiv \lim_{T\to 0} q_{EA}(T)
\; .
\eeq
Equation~(\ref{guess}) can be easily checked experimentally. 

Another related test is to measure the dc susceptibility
\beq
\chi^{\infty}(T,h) = 
\lim_{t_w \rightarrow \infty} \lim_{t \rightarrow \infty} \chi(t,t_w)
\eeq
at constant field $h$, presumably obtainable from 
the field-cooled experimental data,\cite{Eric2} and check if it 
is independent of temperature
everywhere below the (pseudo) Almeida-Thouless line. There is quite a 
bit of evidence in this direction in the literature.\cite{MfcindepT}

One can then extract the curved part of the 
$\chi$ vs $C$ plot of Fig.~\ref{chifig}, that we called
 $f(C)$ from the set of equations
\beqa
C &=& q_d - T y(T)
\; ,
\nn
f(C) &=& y(T)
\; ,
\eeqa
using $T$ as a parameter.

Once (and if) the validity of the approximation is verified, one can 
use the  $\chi$ vs $C$ plot so obtained to check the consistency of the
present reasoning for the cycling experiments.
For example, in the positive cycling experiments 
the difference in TRM 
between
the isothermal and the cycled experiment just after the cycle
should be equal to the difference in height 
in the $\chi$ vs $C$ plot of the intersection of the straight 
lines corresponding
to $T^-$ and $T^+$ (this is best seen by looking at Fig.~\ref{chifig}). 
This non-trivial relation can be checked experimentally.

\section{Conclusions}
\label{section5}

While it is gratifying to be able to reproduce 
the qualitative experimental features analytically, we believe
one should not hasten to declare that this is evidence for the mean-field
scenario holding strictly.
Thus, {\it infinitely} separated time-scales could
in real life be just {\em very} separated scales, the Almeida-Thouless
line could be just a dynamic crossover, etc.

The derivation in this paper is the first 
one done for a microscopic model that on one hand captures the main
features of the experiments and on the other hand demonstrates the difference
in the behaviour of glasses and spin-glasses. Although our calculation 
has relied on a particular model  we believe 
that it will carry through to all classical mean-field spin-glass models.

\vspace{.3cm}

There is however a problem. 
A satisfying scaling of the (isothermal) experimental 
susceptibility and TRM data
{\it simultaneously} was presented in
Ref.~[\raisebox{-.22cm}{\Large \cite{sitges}}]. 
Surprisingly enough, this scaling has only 
two time-regimes as opposed to the many time-scales
that we here invoked
to reproduce analytically the experimental 
results for temperature cyclings within mean-field.
If the scaling used in Ref.~[\raisebox{-.22cm}{\Large \cite{sitges}}] turns out
to hold for all times then the hierarchical
scenario would not apply and some drastic modification 
of the present understanding must be envisaged.
  
\acknowledgements

We thank E. Vincent for providing us with the experimental curves
obtained in Saclay and for several years of very valuable discussions.


\begin{thebibliography}{99}
\bibitem{foot} Experiments in dipolar glasses\cite{Levelut} 
display an ``intermediate'' behaviour in the sense that temperature 
cyclings provoke strong asymmetric results, as in spin-glasses, while 
they also present very strong dependences on the cooling rate, 
a property that is not observed in spin-glasses though is very common
in structural glasses.
\bibitem{foot2} The small reinitialization effects seen in this case
correspond to the re-equilibration of a critical model, and are thus very 
brief.
\bibitem{numerical}
Indeed, we have checked numerically that the full 
$\chi(t)$ is not much affected by the temperature cycling. 
Changing the temperature 
abruptly at $t_1$ one sees a jump in $\chi(t)$ and 
a later quick approach to 
its {\em  temperature-independent}, asymptotic value. This jump is smeared out 
by changing the temperature with a finite velocity, as  
is done in experiments, 
and disappears if the time interval in which one changes the temperature 
is of the order of the FDT regime.
\end{thebibliography}
\end{document}